\documentclass[12pt]{article}
\usepackage{geometry}
\usepackage{makeidx}
\usepackage[T1]{fontenc}
\usepackage[dvipsnames,svgnames,table]{xcolor}
\usepackage{epstopdf}
\usepackage{epsfig}
\usepackage{textcomp}
\usepackage{hyperref}
\usepackage{amsmath}
\usepackage{amssymb}
\usepackage{multirow}
\usepackage{multicol}
\usepackage{booktabs}
\usepackage{lastpage}
\usepackage{hyperref} 
\usepackage{graphicx}
\usepackage{enumerate}
\usepackage{threeparttable}
\usepackage{float}
\usepackage{array}
\usepackage{cite}
\usepackage{color,soul}
\usepackage{tikz}
\makeatletter
\renewcommand\section{\@startsection{section}{1}{\z@}%
                    {-2.5ex \@plus -1ex \@minus -.2ex}%
                    {2.3ex \@plus.2ex}%
                    {\normalfont\large\bfseries}}
\renewcommand\subsection{\@startsection{subsection}{1}{\z@}%
                    {-2.5ex \@plus -1ex \@minus -.2ex}%
                    {2.3ex \@plus.2ex}%
                    {\small\bfseries}}

\begin{document}
\title{\textbf{Applying triangular correlation of angular deviation in muon scattering tomography for multi-block materials via GEANT4 simulations}}
\medskip
\author{\small A. Ilker Topuz$^{1,2}$, Madis Kiisk$^{1,3}$, Andrea Giammanco$^{2}$}
\medskip
\date{\small$^1$Institute of Physics, University of Tartu, W. Ostwaldi 1, 50411, Tartu, Estonia\\
$^2$Centre for Cosmology, Particle Physics and Phenomenology, Universit\'e catholique de Louvain, Chemin du Cyclotron 2, B-1348 Louvain-la-Neuve, Belgium\\
$^3$GScan OU, Maealuse 2/1, 12618 Tallinn, Estonia}
\maketitle
\begin{abstract}
The possibility to exploit the muon scattering for the elemental discrimination of materials in a given volume is well known. When more than one material is present along the muon path, it is often important to discern the order in which they are stacked. The scattering angle due to the target volume can be split into two interior angles in the tomographic setups based on the muon scattering, and we call this property as the triangular correlation where the sum of these two interior angles is equal to the scattering angle. In this study, we apply this triangular correlation for a multi-block material configuration that consist of concrete, stainless steel, and uranium. By changing the order of this material set, we employ the GEANT4 simulations and we show that the triangular correlation is valid in the multi-block material setups, thereby providing the possibility of supportive information for the coarse prediction of the material order in such configurations. 
\end{abstract}
\textbf{\textit{Keywords: }}  Muon Scattering Tomography; Multi-block materials; Scattering Angle; Exterior Angle; Interior Angle; Monte Carlo Simulations; GEANT4
\section{Introduction}
In the muon scattering tomography~\cite{pesente2009first, Checchia_2016, procureur2018muon, bonechi2020atmospheric}, the scattering angle due to the target volumes and its associated statistics serve to discriminate as well as to reconstruct the corresponding volume-of-interests (VOIs) in the image reconstruction techniques such as Point-of-Closest Approach (POCA)~\cite{schultz2003cosmic, bandieramonte2013automated, yu2013preliminary, liu2018muon, yang2019novel,zeng2020principle, liu2021muon}.  As described by the regular tomographic configurations based on the muon scattering~\cite{borozdin2003radiographic}, the complete detection system regularly includes a bottom hodoscope below the VOI in addition to a top hodoscope above the VOI  with multiple detector layers present at each hodoscope~\cite{bandieramonte2013automated, yu2013preliminary, zeng2020principle}. In these tomographic setups based on the muon scattering, the scattering angle is commonly determined by constructing a vector~\cite{carlisle2012multiple, nugent2017multiple, poulson2019application} founded on two hit locations at two distinct detector layers within every hodoscope. 

In another study~\cite{topuz2022unveiling}, we already show that the scattering angle might be split into two interior angles, and these interior angles vary depending on the position of the VOI although the scattering angle almost remains as the same. In this study, we apply the triangular correlation on a three-block material configuration that consists of concrete, stainless steel, and uranium in order to check whether the triangular correlation holds. We perform a number of GEANT4 simulations~\cite{agostinelli2003geant4} by changing the material order within our tomographic scheme~\cite{georgadze2021method} consisting of three plastic scintillators manufactured of polyvinyl toluene and we demonstrate that the triangular correlation is conserved for the multi-block material systems. The current study is organized as follows. In section~\ref{triangular_correlation}, we recall the scattering angle as well as the interior separate angles in accordance with the triangular correlation by depicting over our tomographic configuration, and section~\ref{Simulation_properties_Muography2023} is composed of our simulation schemes in order to explore the triangular correlation as well as the material order. While we show our simulation results in section~\ref{Simulation_results_Muography2023}, we draw our conclusions in section~\ref{Conclusion}.
\section{Triangular correlation}
\label{triangular_correlation}
\begin{figure}[H]
\begin{center}
\includegraphics[width=7.5cm]{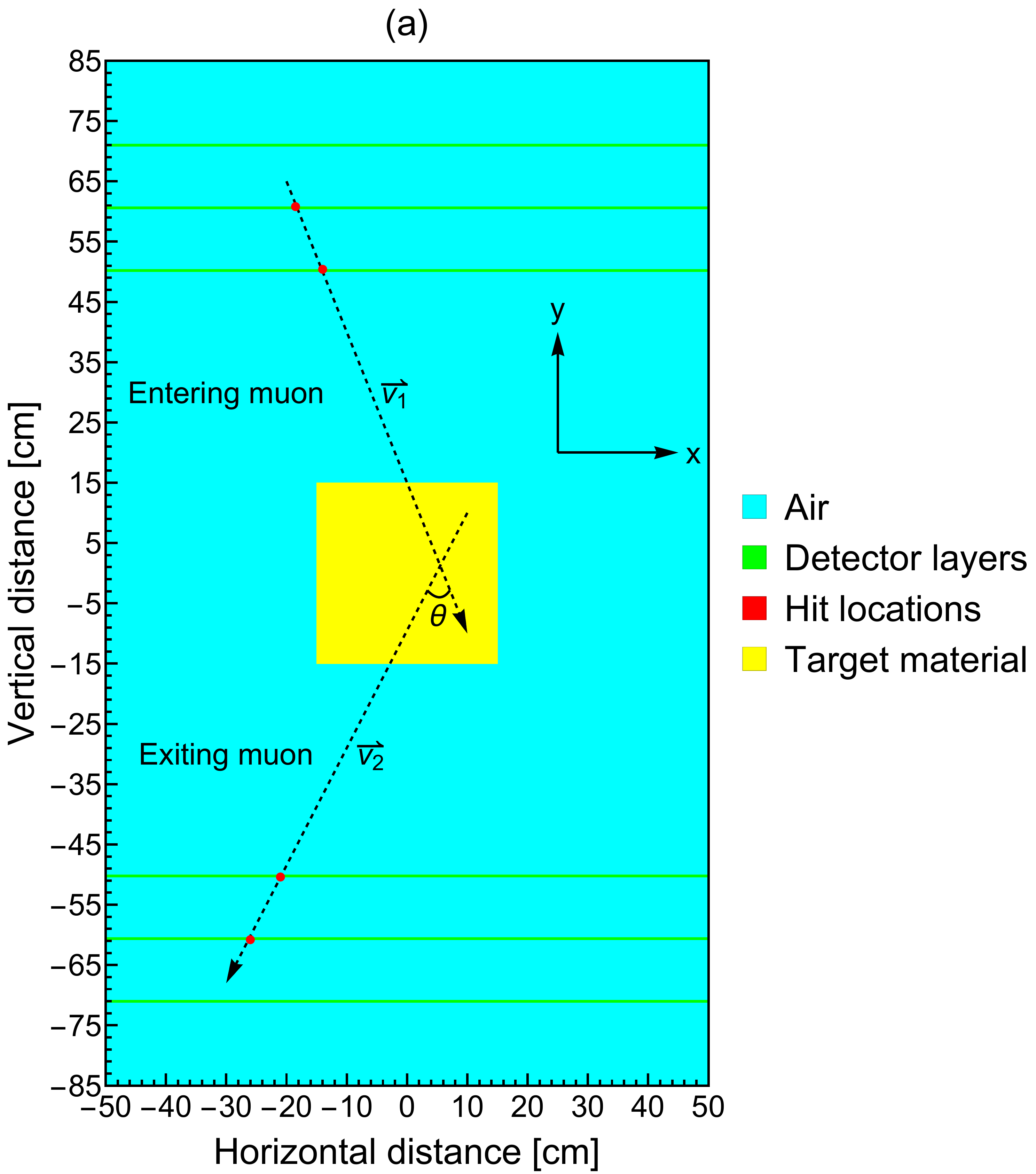}
\includegraphics[width=7.5cm]{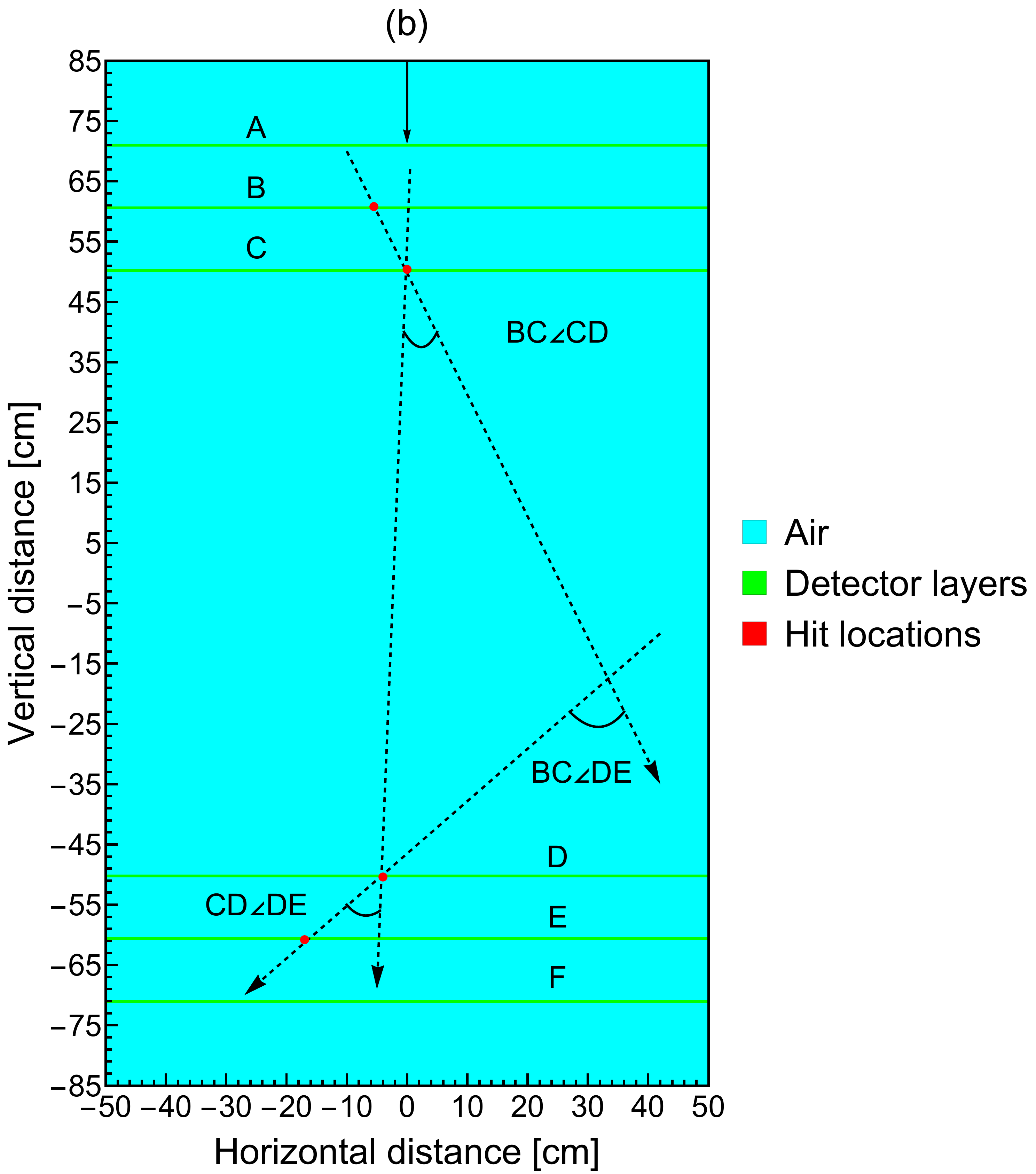}
\caption{Illustration of angular deviation due to the target volume in our tomographic scheme: (a) scattering angle denoted by $\theta$ and (b) triangular correlation between $\theta=\rm BC\angle DE$ and the interior angles denoted by $\rm BC\angle CD$ and $\rm CD\angle DE$ after splitting.}
\label{angular}
\end{center}
\end{figure}

Initially, our tomographic setup is illustrated in Fig.~\ref{angular} (a) where the scattering angle indicated by $\theta$ is determined by building a vector at each section, the components of which are obtained through the hit locations on two detector layers. The scattering angle might be split into two opposite angles by setting up a triangular correlation as delineated Fig.~\ref{angular} (b) where the exterior angle referred to the scattering angle is equal to the superposition of the two non-adjacent angles.
\label{triangular correlation}
\vskip -0.25cm
By recalling that the capital letters listed as $\rm A, B, C, D, E,$ and $\rm F$ in Fig.~\ref{angular} (b) point to the hit locations in the specific detector layers, the conventional scattering angle denoted by $\theta$ that also refers to the exterior angle is commonly expressed as written in~\cite{carlisle2012multiple, nugent2017multiple, poulson2019application}
\begin{equation}
\theta=\rm BC\angle DE=\rm BC\angle CD + \rm CD\angle DE=\arccos\left(\frac{\overrightarrow{\rm BC}\cdot\overrightarrow{\rm DE}}{\left|BC\right|\left|DE\right|}\right)
\label{superpose}
\end{equation}
The same set of four hit locations also gives access to compute two opposite interior angles as defined in 
\begin{equation}
\rm BC\angle CD=\arccos\left(\frac{\overrightarrow{\rm  BC}\cdot\overrightarrow{\rm  CD}}{\left|BC\right|\left|CD\right|}\right)
\end{equation}
and
\begin{equation}
\rm CD\angle DE=\arccos\left(\frac{\overrightarrow{\rm  CD}\cdot\overrightarrow{\rm  DE}}{\left|CD\right|\left|DE\right|}\right)
\end{equation}
The average angular deviation of any combination, i.e  $\overline{x\angle y}$, at a given energy value is determined by averaging over $N$ number of the non-absorbed/non-decayed muons as defined in 
\begin{equation}
\overline{x\angle y}=\frac{1}{N}\sum^{N}_{i=1}(x\angle y)_{i}
\end{equation}
\section{Simulation properties}
\label{Simulation_properties_Muography2023}
\begin{figure}[H]
\begin{center}
\includegraphics[width=10cm]{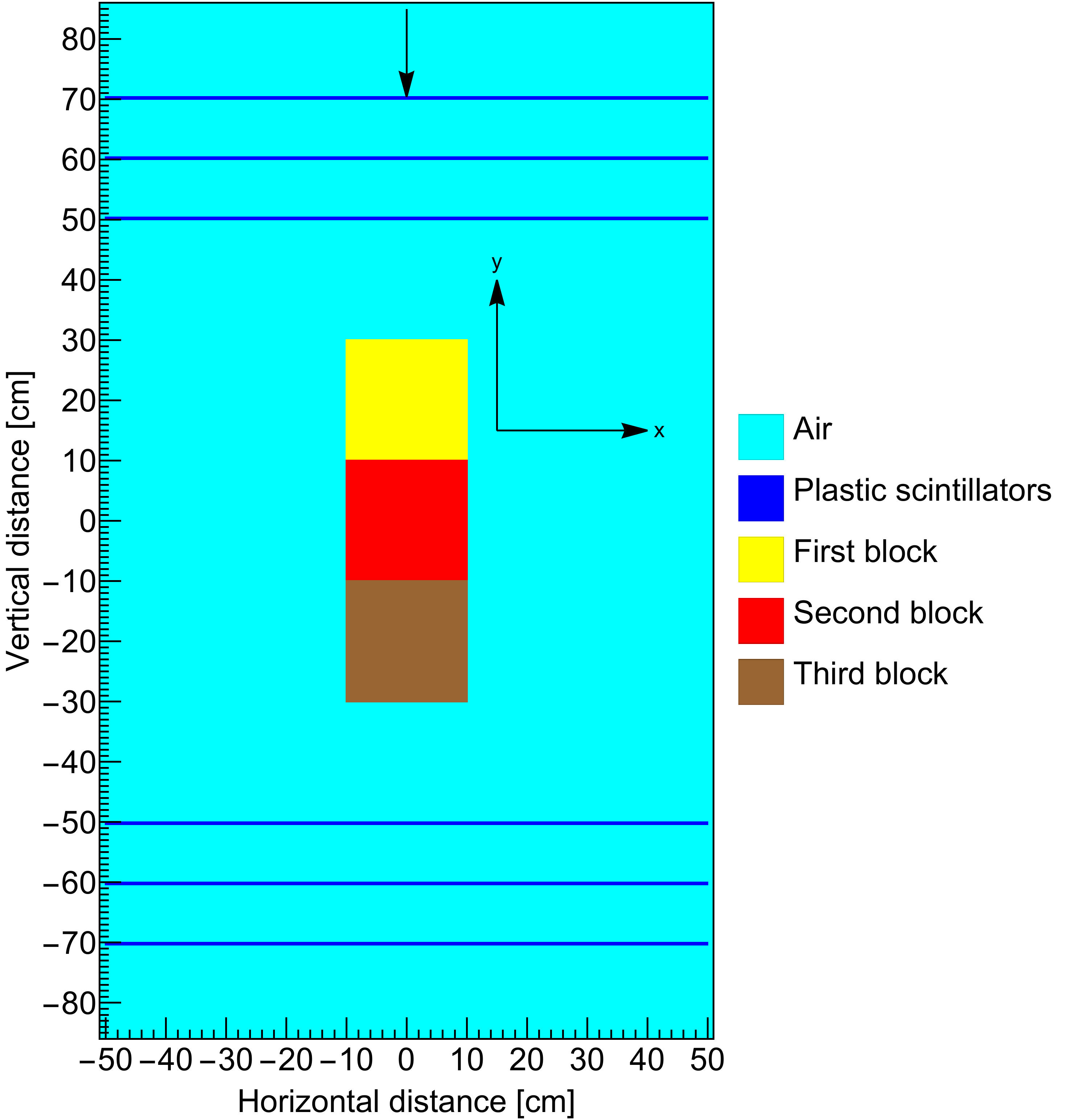}
\caption{Simulation setup for the multi-block material configuration that consists of concrete, stainless steel, and uranium.}
\label{Simulation_setup_Muography2023}
\end{center}
\end{figure}

Following the definition of the triangular correlation and the associated angles of this correlation collected based on the tracked hits from the detector layers, we perform a series of GEANT4 simulations in order to verify the triangular correlation. Since we have three different blocks, we define six combinations that consist of concrete+stainless steel+uranium (case I), concrete+uranium+stainless steel (case II), stainless steel+concrete+uranium (case III), stainless steel+uranium+concrete (case IV), uranium+concrete+stainless steel (case V), and uranium+stainless steel+concrete (case VI) according to Fig.~\ref{Simulation_setup_Muography2023} . Apart from the material order, each block is a cubic volume with the dimensions of $20\times20\times20$ $\rm cm^{3}$.
To briefly summarize, our tomographic setup in GEANT4 simulations is composed of three plastic scintillators made out of polyvinyl toluene with the dimensions of $100\times0.4\times100$ $\rm cm^{3}$ at every section. We utilize a central mono-directional uniform muon beam as indicated by a downward black arrow in Fig.~\ref{Simulation_setup_Muography2023}, and the uniform energy distribution~\cite{anghel2015plastic} lies on an interval between 0.1 and 8 GeV for the reason of more favorable numerical stability. Since the current aperture of the entire detection geometry commonly only accepts the narrow angles apart from the very rare entries around the corners, this beam setup is considered significantly reliable by reminding that the distribution of the incident angle $(\alpha)$ approximately corresponds to $\rm cos^{2}(\alpha)$ for an interval between $-\pi/2$ and  $\pi/2$~\cite{yanez2021method}. The number of the simulated muons in each defined position is $10^{5}$. The tomographic components in the GEANT4 simulations are defined in agreement with the G4/NIST database, and the preferred physics list is FTFP$\_$BERT. The simulation features are listed in Table~\ref{Simulation_features_Muography2023}.
\begin{table}[H]
\begin{center}
\begin{footnotesize}
\caption{Simulation properties.}
\begin{tabular}{cc}
\toprule
\toprule
Particle & $\mbox{\textmu}^{-}$\\
Beam direction & Vertical\\
Momentum direction & (0, -1, 0)\\
Source geometry & Planar\\
Initial position (cm) & ([-0.5, 0.5], 85, [-0.5, 0.5])\\
Number of particles & $10^{5}$\\
Energy interval (GeV) & [0, 8]\\
Energy cut-off (GeV) & 0.1\\
Bin step length (GeV) & 0.5\\
Energy distribution & Uniform\\
Material database & G4/NIST\\
Reference physics list & FTFP$\_$BERT\\
\bottomrule
\bottomrule
\label{Simulation_features_Muography2023}
\end{tabular}
\end{footnotesize}
\end{center}
\end{table}
The muon tracking is accomplished by G4Step, and the tracked hit locations are post-processed by the support of a Python script where the scattering angle and the interior non-adjacent angles are initially computed for every single non-absorbed/non-decayed muon, then the uniform energy spectrum limited by 0.1 and 8 GeV is divided into 16 bins by marching with a step of 0.5 GeV, and each obtained energy bin is labeled with the central point in the energy sub-interval. Finally, the determined angles are averaged for the associated energy bins.
\section{Simulation outcomes}
\begin{figure}[H]
\begin{center}
\includegraphics[width=8cm]{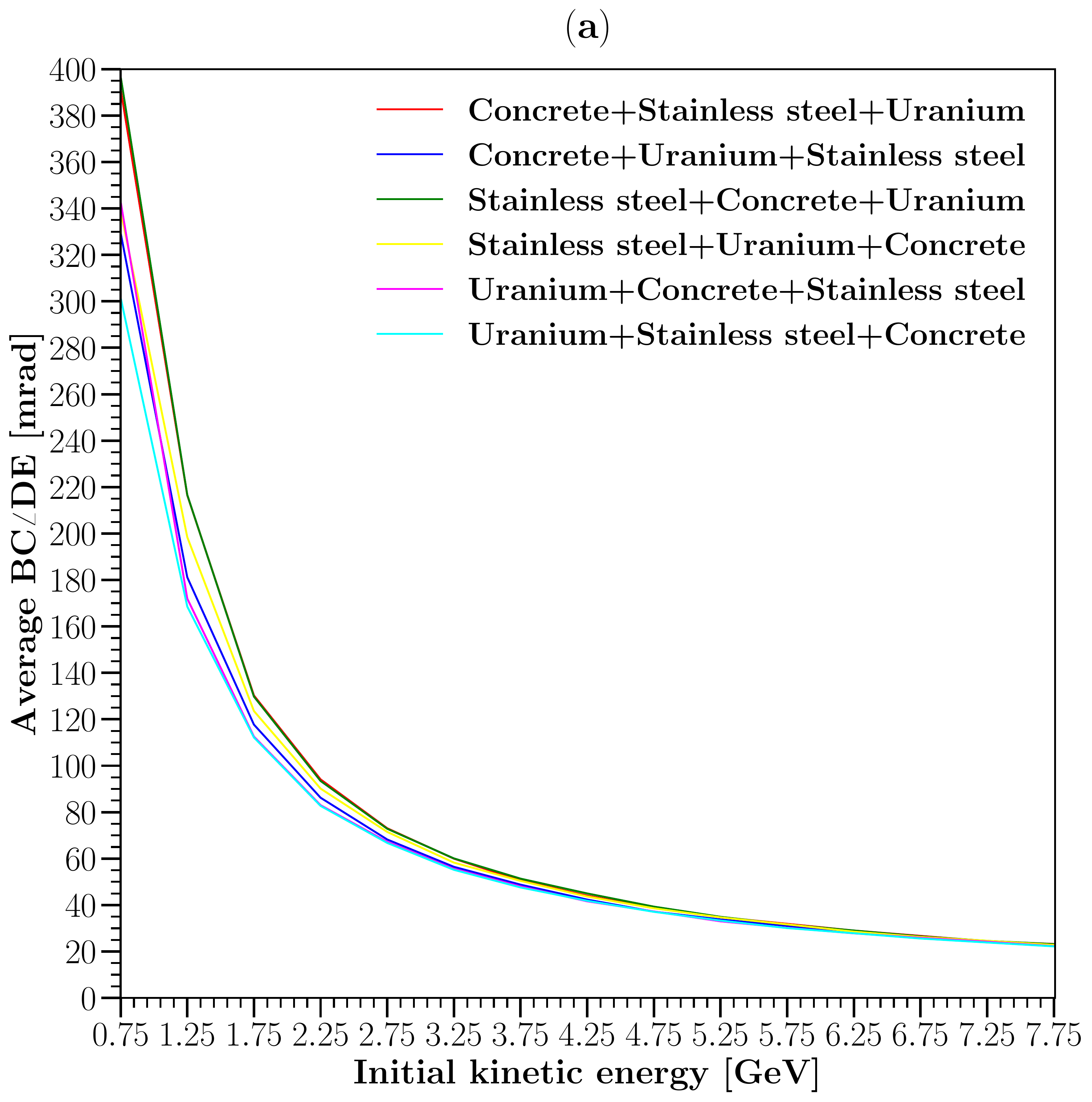}
\includegraphics[width=8cm]{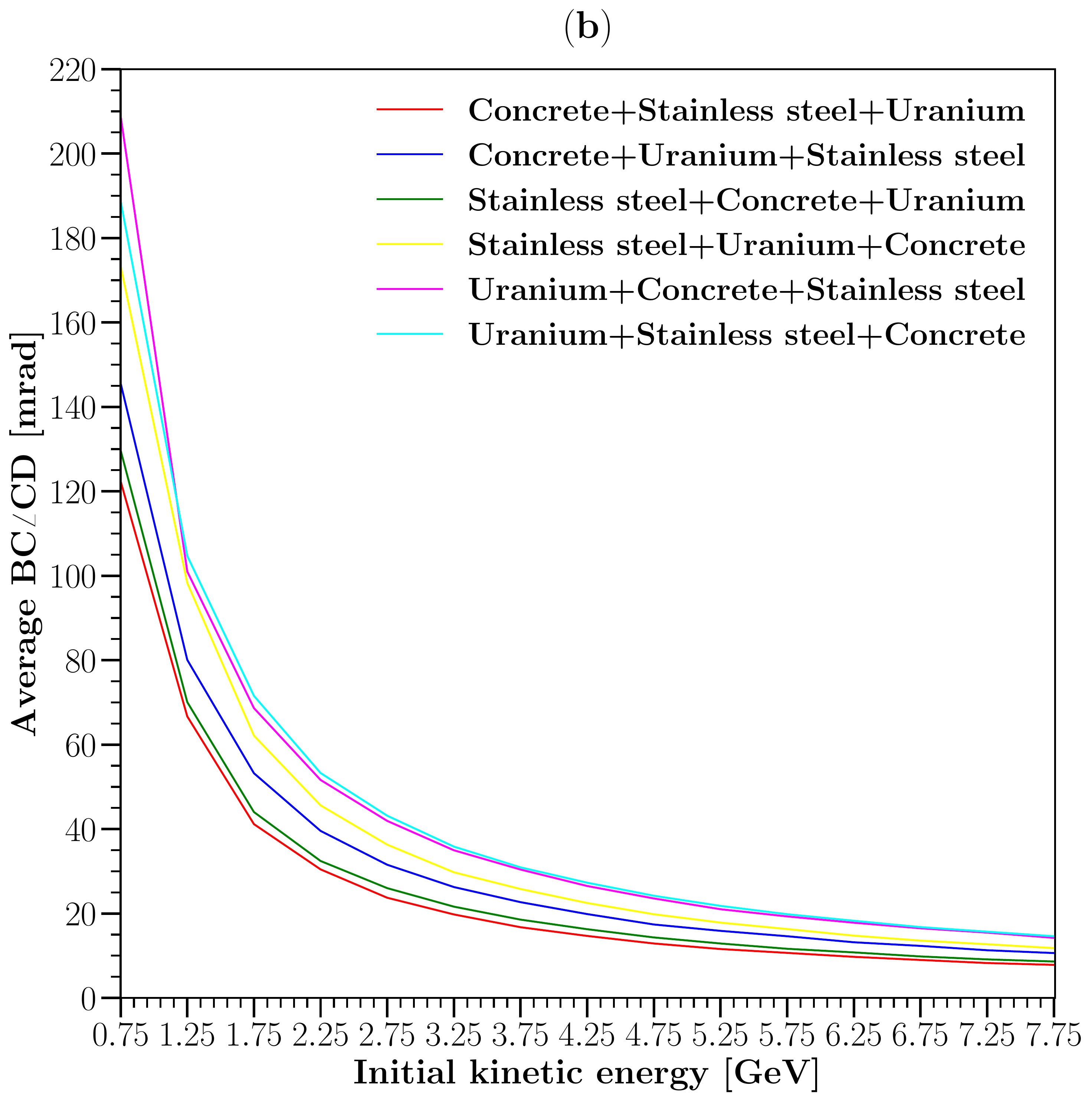}
\includegraphics[width=8cm]{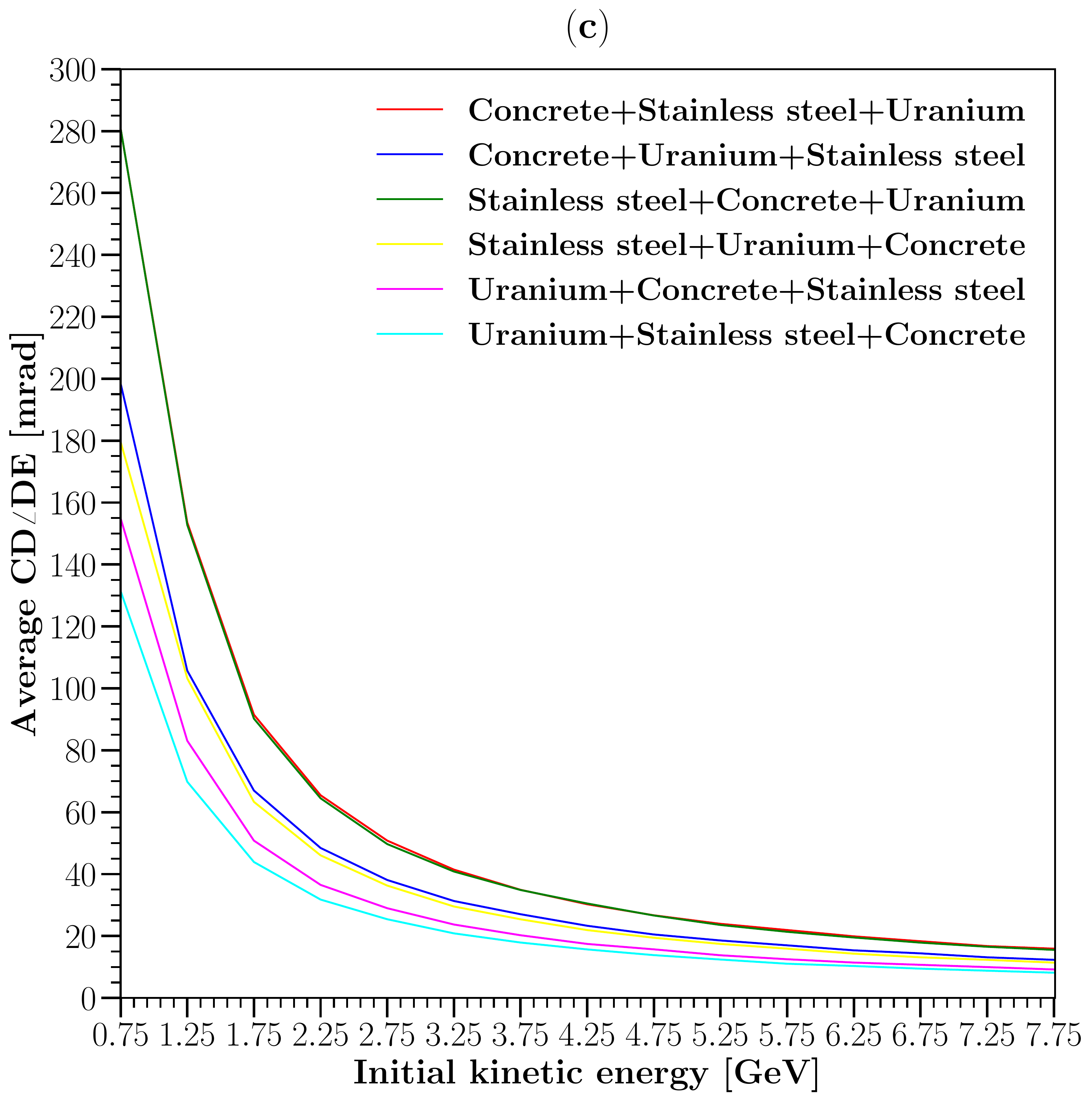}
\includegraphics[width=8cm]{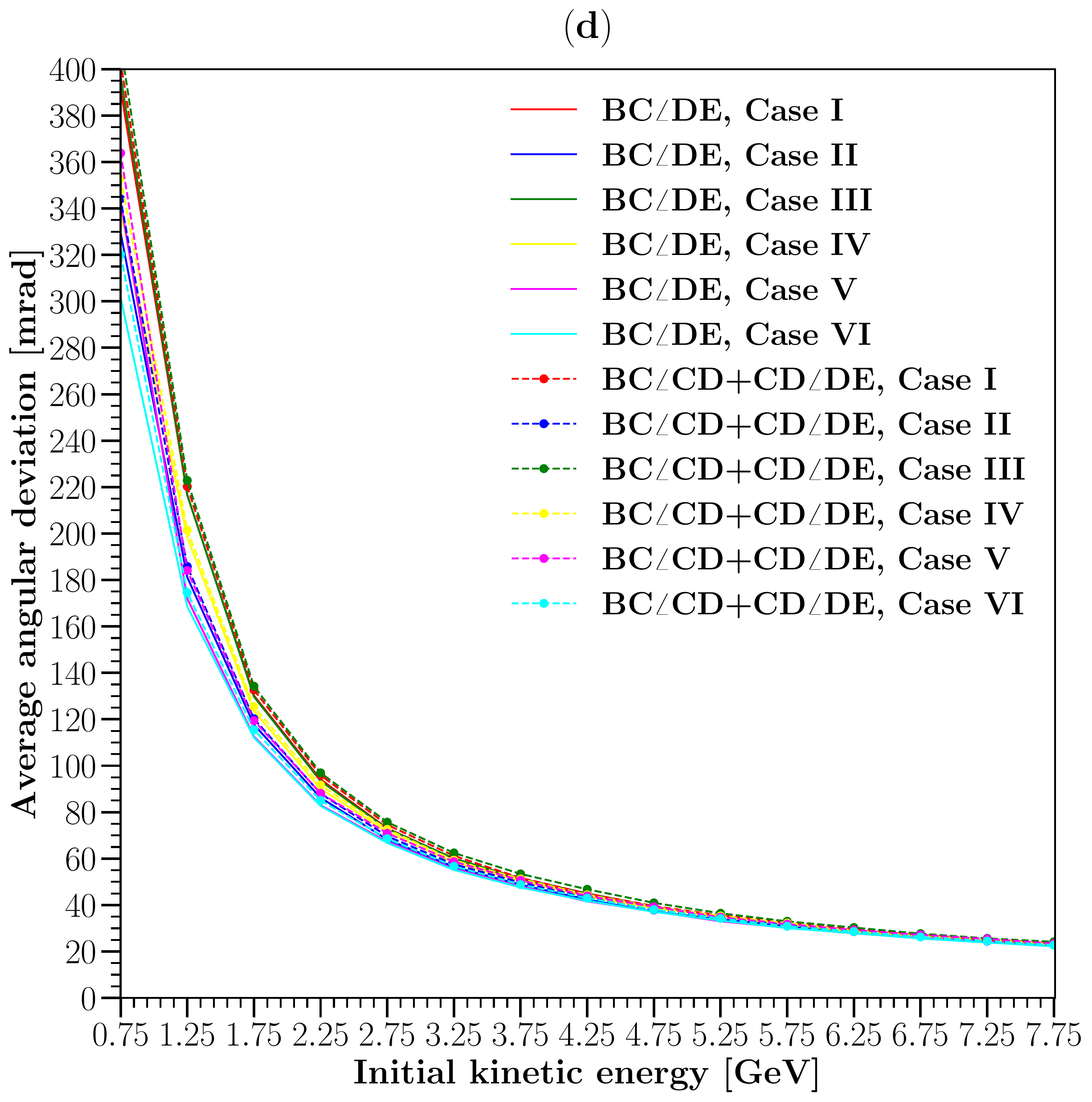}
\caption{Simulation outcomes for the multi-block material system that consists of concrete, stainless steel, and uranium.}
\label{multiblock_angles}
\end{center}
\end{figure}

We first investigate the average $\rm BC\angle DE$ as shown in Fig.~\ref{multiblock_angles} (a). For the average $\rm BC\angle DE$, we show that the angular values converge in the high energy bins. According to Figs.~\ref{multiblock_angles} (b) and (c), this convergence is not observed for the average $\rm BC\angle CD$ as well as the average $\rm CD\angle DE$. Moreover, the average $\rm BC\angle CD$ and the average $\rm CD\angle DE$ result in more distinct curves compared to the average $\rm BC\angle DE$, thereby implying an opportunity to coarsely predict the material order in the multi-block material systems. Finally in Fig.~\ref{multiblock_angles} (d), by summing up the average $\rm BC\angle CD$ and the average $\rm CD\angle DE$, we demonstrate that Eq.~(\ref{superpose}) holds for the multi-block materials. 
\label{Simulation_results_Muography2023}
\section{Conclusion}
\label{Conclusion}
In conclusion, we show that the triangular correlation of the angular deviation holds for the multi-block material configurations. Moreover, we also imply that this triangular correlation might provide supportive information for the coarse prediction of the material order in such configurations.
\bibliographystyle{elsarticle-num}
\nocite{*}
\bibliography{multi_block.bib, triangular.bib} 
\end{document}